\documentclass[twocolumn,showpacs,preprintnumbers,amsmath,amssymb]{revtex4}

\usepackage{graphicx}
\usepackage{epsfig}
\usepackage{rotating}
\usepackage{dcolumn}
\usepackage{bm}

\begin{document}

\preprint{HEP/123-qed}
\title{A narrow structure in the excitation function of $\eta$-photoproduction off the neutron}
\author{
  D.~Werthm\"uller$^1$,
  L.~Witthauer$^1$, 
  I.~Keshelashvili$^{1}$, 
  P.~Aguar-Bartolom${\acute{\rm e}}$$^{2}$,   
  J.~Ahrens$^{2}$,
  J.R.M.~Annand$^{3}$,
  H.J.~Arends$^{2}$,
  K.~Bantawa$^{4}$,
  R.~Beck$^{2,5}$,
  V.~Bekrenev$^{6}$,
  A.~Braghieri$^{7}$,
  D.~Branford$^{8}$,
  W.J.~Briscoe$^{9}$,
  J.~Brudvik$^{10}$,
  S.~Cherepnya$^{11}$,
  B.~Demissie$^{9}$,
  M.~Dieterle$^{1}$,  
  E.J.~Downie$^{2,3,9}$,
  P.~Drexler$^{12}$,
  L.V. Fil'kov$^{11}$, 
  A. Fix$^{13}$,      
  D.I.~Glazier$^{8}$,
  D. Hamilton$^{3}$,
  E.~Heid$^{2}$,
  D.~Hornidge$^{14}$,
  D.~Howdle$^{3}$,
  G.M.~Huber$^{15}$,
  I.~Jaegle$^{1}$,
  O.~Jahn$^{2}$,
  T.C.~Jude$^{8}$,
  A.~K{\"a}ser$^{1}$,   
  V.L.~Kashevarov$^{2,11}$,
  R.~Kondratiev$^{16}$,
  M.~Korolija$^{17}$,
  S.P.~Kruglov$^{6}$, 
  B.~Krusche$^1$, 
  A.~Kulbardis$^{6}$,  
  V.~Lisin$^{16}$,
  K.~Livingston$^{3}$,
  I.J.D.~MacGregor$^{3}$,
  Y.~Maghrbi$^{1}$,
  J.~Mancell$^{3}$, 
  D.M.~Manley$^{4}$,
  Z.~Marinides$^{9}$,
  M.~Martinez$^{2}$,
  J.C.~McGeorge$^{3}$,
  E.F.~McNicoll$^{3}$,
  V.~Metag$^{12}$,
  D.G.~Middleton$^{14}$,
  A.~Mushkarenkov$^{7}$,
  B.M.K.~Nefkens$^{10}$,
  A.~Nikolaev$^{2,5}$,
  R.~Novotny$^{12}$,
  M.~Oberle$^{1}$,
  M.~Ostrick$^{2}$,
  B.~Oussena$^{2,9}$, 
  P.~Pedroni$^{7}$,
  F.~Pheron$^{1}$,
  A.~Polonski$^{16}$,
  S.N.~Prakhov$^{2,9,10}$,
  J.~Robinson$^{3}$,   
  G.~Rosner$^{3}$,
  T.~Rostomyan$^{1}$,
  S.~Schumann$^{2,5}$,
  M.H.~Sikora$^{8}$,
  D.~Sober$^{18}$,
  A.~Starostin$^{10}$,
  I.~Supek$^{17}$,
  M.~Thiel$^{2,12}$,
  A.~Thomas$^{2}$,
  M.~Unverzagt$^{2,5}$,
  D.P.~Watts$^{8}$\\
(Crystal Ball/TAPS experiment at MAMI, the A2 Collaboration)
}
\affiliation{
  $^{1}$\mbox{Departement f\"ur Physik, Universit\"at Basel, Switzerland}\\
  $^{2}$\mbox{Institut f\"ur Kernphysik, Universit\"at Mainz, Germany}\\
  $^{3}$\mbox{SUPA, School of Physics and Astronomy, University of Glasgow, Glasgow G12 8QQ, UK}\\
  $^{4}$\mbox{Kent State University, Kent, OH, USA}\\  
  $^{5}$\mbox{Helmholtz-Institut f\"ur Strahlen- und Kernphysik, Universit\"at Bonn, Germany}\\
  $^{6}$\mbox{Petersburg Nuclear Physics Institute, Gatchina, Russia}\\
  $^{7}$\mbox{INFN Sezione di Pavia, Pavia, Italy}\\
  $^{8}$\mbox{SUPA, School of Physics, University of Edinburgh, Edinburgh EH9 3JZ, UK}\\
  $^{9}$\mbox{Institute for Nuclear Studies, The George Washington University, Washington, DC, USA}\\
  $^{10}$\mbox{University of California at Los Angeles, Los Angeles, CA, USA}\\
  $^{11}$\mbox{Lebedev Physical Institute, Moscow, Russia}\\
  $^{12}$\mbox{II. Physikalisches Institut, Universit\"at Giessen, Germany}\\
  $^{13}$\mbox{Laboratory of Mathematical Physics, Tomsk Polytechnic University, Tomsk, Russia}\\
  $^{14}$\mbox{Mount Allison University, Sackville, NB E4L 1E6, Canada}\\
  $^{15}$\mbox{University of Regina, Regina, SK S4S 0A2, Canada}\\
  $^{16}$\mbox{Institute for Nuclear Research, Moscow, Russia}\\
  $^{17}$\mbox{Rudjer Boskovic Institute, Zagreb, Croatia}\\
  $^{18}$\mbox{The Catholic University of America, Washington, DC, USA}\\
}
\date{\today}

\begin{abstract}
The photoproduction of $\eta$-mesons off nucleons bound in $^2$H and $^3$He 
has been measured in coincidence with recoil protons and recoil neutrons for incident 
photon energies from threshold up to 1.4~GeV. The experiments were performed at 
the Mainz MAMI accelerator, using the Glasgow tagged photon facility. Decay photons
from the $\eta\rightarrow 2\gamma$ and $\eta\rightarrow 3\pi^0$ decays and the recoil 
nucleons were detected with an almost $4\pi$ electromagnetic calorimeter 
combining the Crystal Ball and TAPS detectors. The data from both targets are of excellent 
statistical quality and show a narrow structure in the excitation function of 
$\gamma n\rightarrow n\eta$. The results from the two measurements are consistent taking 
into account the expected effects from nuclear Fermi motion. The best estimates for position 
and intrinsic width of the structure are $W$ = (1670$\pm$5)~MeV and $\Gamma$~=~(30$\pm$15)~MeV. 
For the first time precise results for the angular dependence of this structure have 
been extracted. 
\end{abstract}

\maketitle

Photo- and electroproduction of mesons has become a primary tool for the investigation of 
the excitation spectrum of the nucleon \cite{Krusche_03,Burkert_04,Klempt_10,Aznauryan_12}. 
So far, most efforts have been devoted to the excitation spectrum of the proton, 
simply because free neutron targets are not available. However, since the electromagnetic 
excitations are isospin dependent, such measurements are indispensable. Experiments 
therefore have to make use of quasi-free neutrons bound in light nuclei, in particular 
in the deuteron. The specific problems of using quasi-free neutron targets have been 
studied in detail during the last few years \cite{Jaegle_08,Jaegle_11a,Jaegle_11b,Krusche_11}.

An exciting result was a narrow structure in the excitation function of 
$\eta$-photoproduction off the neutron, which was first reported from the GRAAL 
experiment in Grenoble \cite{Kuznetsov_07} and subsequently seen in measurements 
at ELSA in Bonn \cite{Jaegle_08,Jaegle_11a}, and at LNS in Sendai \cite{Miyahara_07}. 
The study of $\eta$-photoproduction off the neutron was motivated by several unresolved 
issues. Prior to the above mentioned experiments, $\eta$-photoproduction off the 
deuteron (or other light nuclear targets) had been studied with incident photon 
energies below 1~GeV \cite{Krusche_95a,Hoffmann_97,Hejny_99,Weiss_01,Weiss_03}. 
There, it is dominated by the excitation of the S$_{11}$(1535) resonance 
\cite{Krusche_95,Krusche_97} (see \cite{Krusche_03} for a summary). 
However, reaction models like the $\eta$-MAID model \cite{Chiang_02} predicted a 
rapid change of the neutron/proton cross section ratio at higher incident-photon 
energies. The electromagnetic excitation of the D$_{15}$(1675) state is Moorhouse 
suppressed \cite{Moorhouse_66} for the proton and expected to contribute much more
strongly for the neutron. Related structures should correspond to typical parameters 
of nucleon resonances in the 1.5~GeV - 2~GeV mass range, i.e. to widths greater 
than 100~MeV. There were also predictions for a narrow structure
related to the conjectured baryon anti-decuplet \cite{Diakonov_97}.
Taking together the results from [20,21,22] the non-strange P$_{11}$-like
member of the anti-decuplet should be electromagnetically excited more 
strongly on the neutron, should have a large decay branching ratio to N$\eta$, 
an invariant mass around 1.7 GeV, and a width of a few tens of MeV.
Surprisingly, all experiments which tried to identify a corresponding 
structure in the $\gamma n\rightarrow n\eta$ reaction reported a positive result 
\cite{Kuznetsov_07,Miyahara_07,Jaegle_08,Jaegle_11a}. Recently, evidence for this
structure was also claimed for the $\gamma n\rightarrow n\gamma '$ reaction 
\cite{Kuznetsov_11}. The Review of Particle Physics \cite{PDG} lists the results as 
tentative evidence for a one-star isospin $I=1/2$ nucleon resonance close to 
1.68~GeV with narrow width and otherwise unknown properties. 

There are two issues that need urgent clarification: how robust is the 
experimental evidence for this narrow structure and, if it is confirmed, what 
is its nature? The present Letter reports results from high statistics measurements 
of quasi-free $\eta$-photoproduction off nucleons bound in the deuteron and in 
$^3$He nuclei, which establish the structure beyond any doubt and reveal its
angular distribution. The experiments were performed at the tagged photon beam 
\cite{McGeorge_08} of the Mainz MAMI accelerator \cite{Walcher_90} with liquid 
deuterium and liquid $^3$He targets. For the deuterium three different beam times 
with varying parameters for the target (length between 3.02 cm and 4.72 cm; surface 
densities 0.147 - 0.230 nuclei/barn) and different trigger conditions were analyzed. 
For the helium measurement the target length (surface density) was 5.08 cm 
(0.073 nuclei/barn). Electron beam energies were between 1508~MeV and 1557~MeV.   

The decay photons from the $\eta$-mesons and the recoil nucleons from quasi-free 
production reactions were detected with an electromagnetic calorimeter, combining 
the Crystal Ball (CB) \cite{Starostin_01} and TAPS \cite{Gabler_94} 
detectors in a setup covering almost the full solid angle ($\approx$ 98\% of $4\pi$). 
Charged-particle identification was provided by additional scintillation
detectors around the target (PID) \cite{Watts_04} and in front of the TAPS wall (CPV). 
More details are given in Ref. \cite{Pheron_12}.

The separation of photons, protons, and neutrons in the TAPS detector used 
the signals from the plastic scintillators, a time-of-flight (ToF) versus energy analysis, 
and a pulse-shape analysis (PSA) for the BaF$_2$ modules. 
In the CB, the PID allowed photons and neutrons to be separated from charged particles and
protons to be distinguished from charged pions using a $\Delta E - E$ analysis.
More details are given in \cite{Zehr_12}. Photons and neutrons could not be 
distinguished in the CB. Neutral CB hits were taken as candidates for both.
Events with $\eta$-mesons in coincidence with either recoil protons or recoil neutrons 
were analyzed. The reaction identification used standard invariant and missing mass 
techniques and coplanarity between the $\eta$-meson and the recoil nucleon, as in 
\cite{Jaegle_11a}. For both nuclei $\eta$-mesons were identified from their 
$\eta\rightarrow \gamma\gamma$ and $\eta\rightarrow 3\pi^0\rightarrow 6\gamma$ decays.
For events with three ($\eta\rightarrow 2\gamma$ decay) or seven 
($\eta\rightarrow 3\pi^0$ decay) neutral hits, a $\chi^2$ test on the invariant mass 
of photon pairs, compared to the $\eta$-mass or to three $\pi^0$ invariant masses was 
used to identify the most probable assignment of the neutron candidate. For hits in 
TAPS those candidates had also to pass the PSA and ToF-versus-energy filters. 

\begin{figure}[htb]
\centerline{\resizebox{0.46\textwidth}{!}{%
  \includegraphics{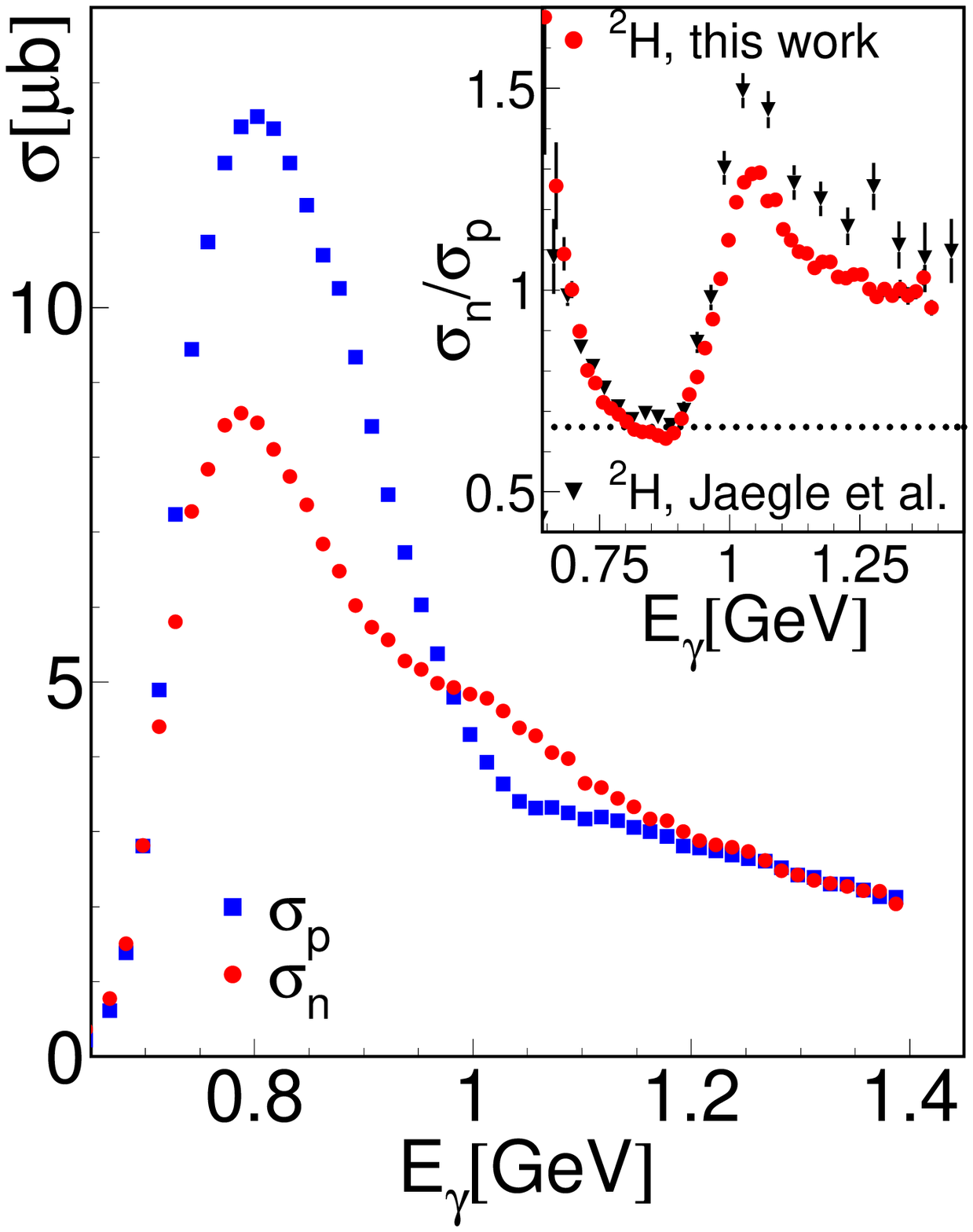}
  \includegraphics{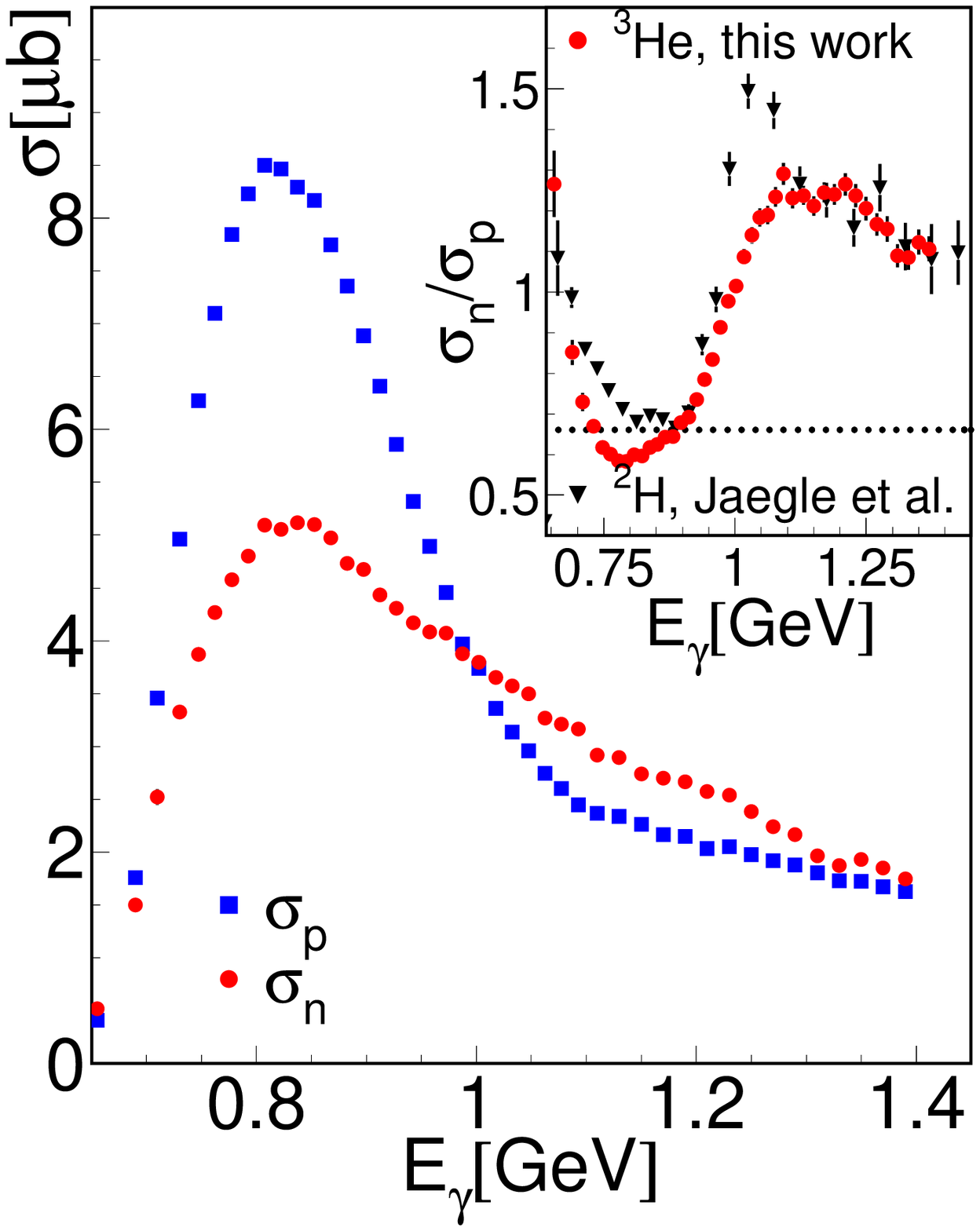}
}}    
\centerline{\resizebox{0.46\textwidth}{!}{%
  \includegraphics{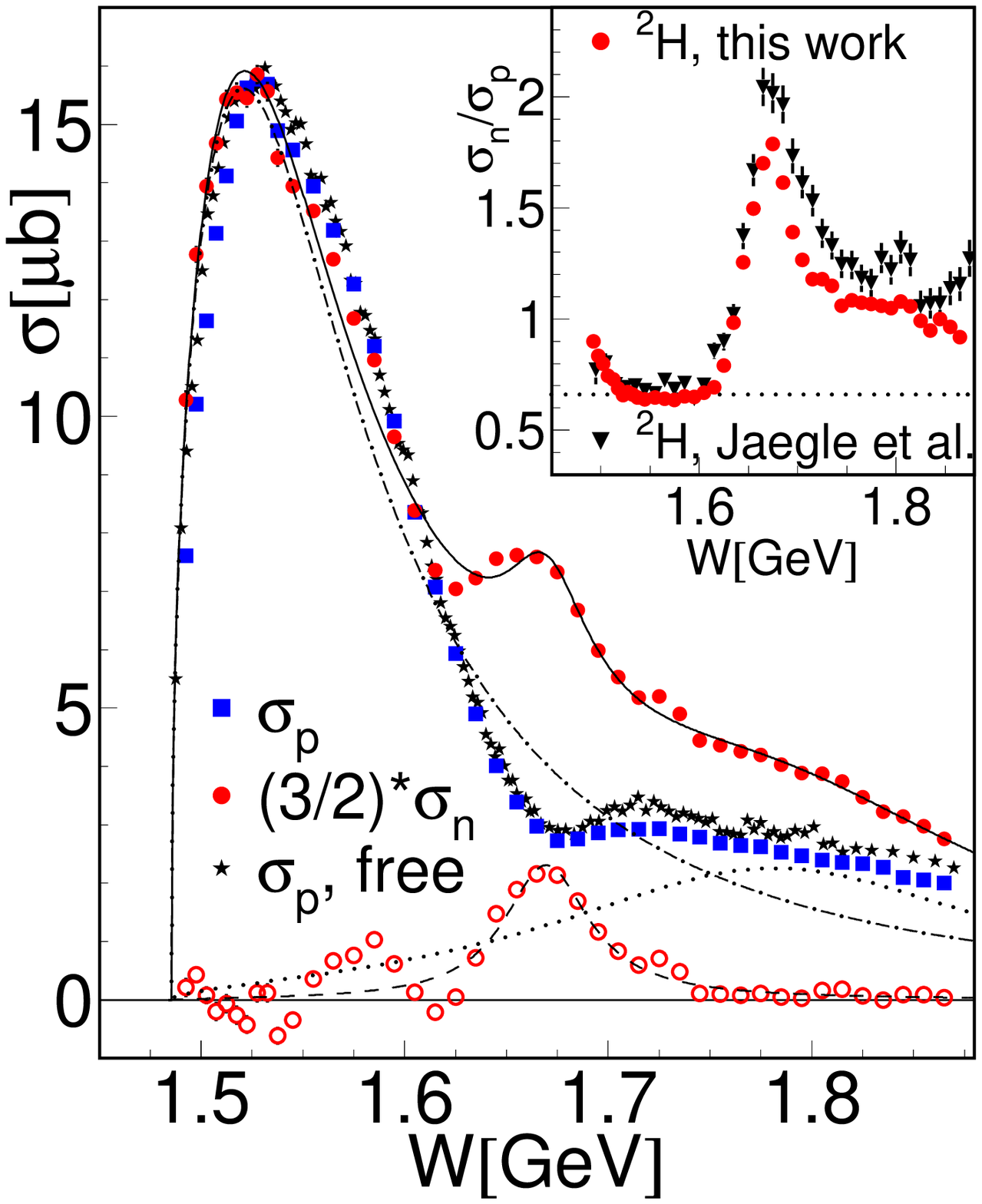}
  \includegraphics{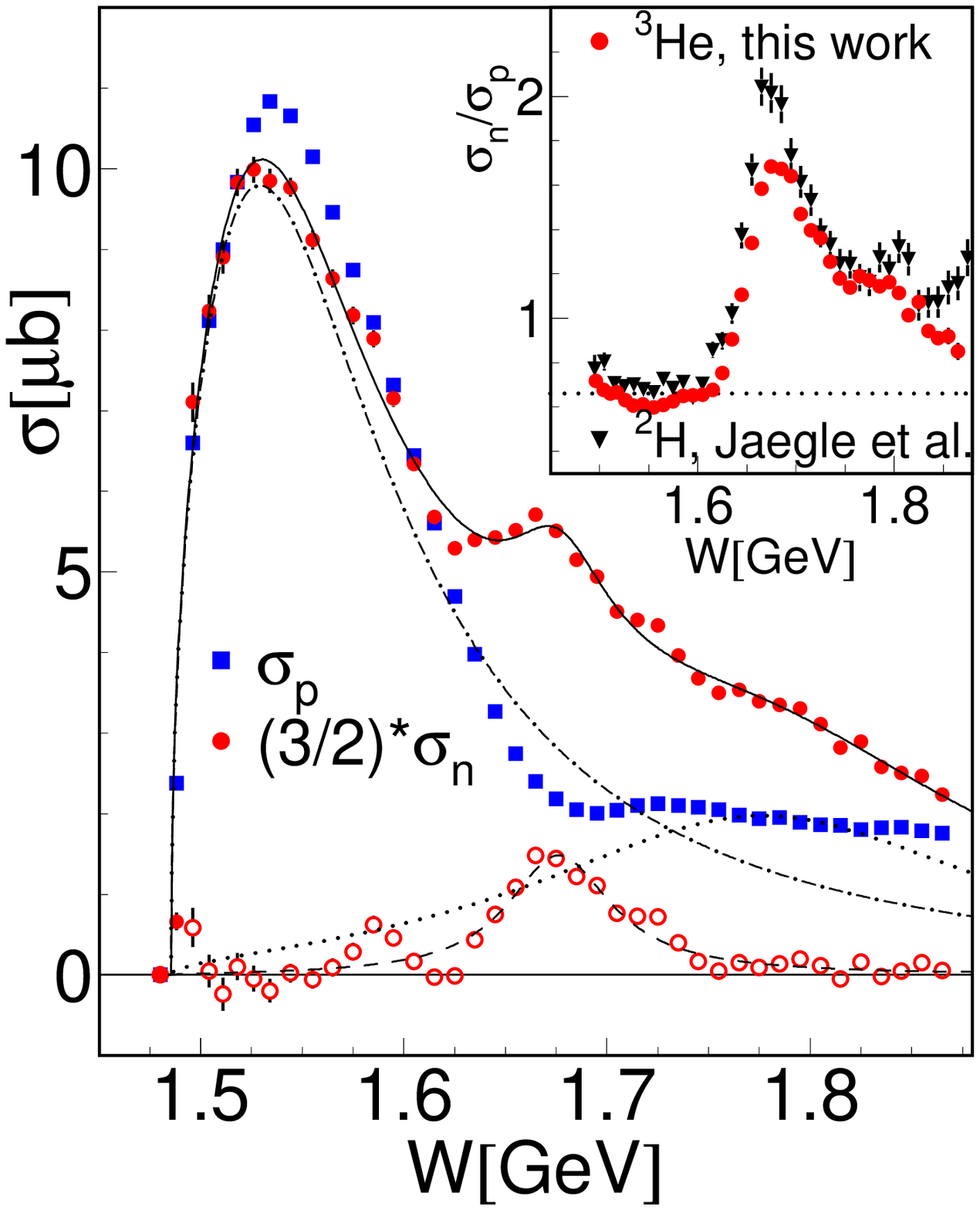}  
}}

\caption{Top: Total cross sections $\sigma_p$ (coincident protons, blue squares) 
and $\sigma_n$ (coincident neutrons, red circles) as function of incident photon energy 
$E_{\gamma}$. Left-hand side: deuterium target. Right-hand side: helium target. 
Bottom: same as function of reconstructed $\eta N$ invariant mass $W$.
Black stars: results for free proton \cite{McNicoll_10}. The open red circles are present
data after subtraction of the fitted S$_{11}$ and background components. 
Curves: fit results for S$_{11}$ resonance (dash-dotted), background (dotted), 
narrow structure (dashed), and full fit (solid).
Inserts for all figures: $\sigma_n/\sigma_p$ ratio from present work (red circles) and from 
Ref.~\cite{Jaegle_11a} (black triangles).
}
\label{fig:total}       
\end{figure}

Total cross sections as function of incident photon energy are shown in 
Fig.~\ref{fig:total}, upper part. They have been averaged over the two $\eta$-decay
modes. For both nuclei the $\sigma_n/\sigma_p$ ratio shows a rapid rise around 1~GeV. 
As expected, the structure has a slow rise for the $^3$He target due to the larger 
Fermi momenta.    

The effects of Fermi motion can be removed when instead of the incident photon energy, 
the invariant mass $W$ of the nucleon - meson pair in the final state 
is used. In case of the three-body $np\eta$ final state for the deuteron, $W$  
can be reconstructed from the measured four-momentum of the $\eta$-meson, 
the energy of the incident photon, and the direction of the recoil nucleon 
\cite{Jaegle_11a}.
For the reaction $\gamma ^3\mbox{He}\rightarrow \eta X$, reconstruction is only
possible under the assumption of vanishing relative momentum between the two
spectator nucleons. This approximation may result in a poorer resolution for the
reconstruction of $W$, however, the effect does not appear to be large. 
The results of this analysis are summarized in the bottom part of Fig.~\ref{fig:total}. 
The proton data measured with the deuterium target are compared to the cross section
for the free $\gamma p\rightarrow p\eta$ reaction from Ref.~\cite{McNicoll_10}. 
The agreement is excellent, and demonstrates the validity of the reconstruction method.  

\begin{figure}[htb]
\centerline{
\resizebox{0.46\textwidth}{!}{%
  \includegraphics{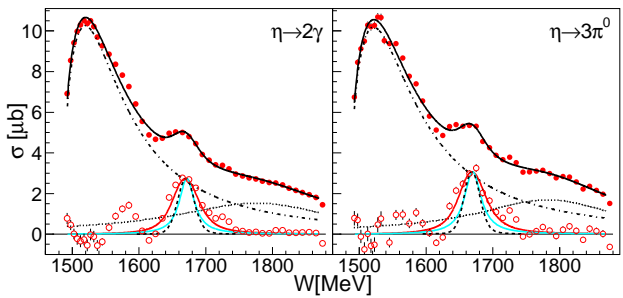}
}}
\caption{Excitation functions for $\gamma n\rightarrow n\eta$ from the deuteron target.
Left-hand side from $\eta\rightarrow 2\gamma$ decay, right-hand side from 
$\eta\rightarrow 6\gamma$ decay. Solid curve and dashed curves: fit components like in 
Fig.~\ref{fig:total}. Open symbols: data with background fit subtracted (scaled up by
factor of 2). Curves at bottom: experimental resolution dashed (black), fitted signal (red), 
intrinsic signal shape (light blue) (see text).
}
\label{fig:dexi}       
\end{figure}

\begin{figure}[htb]
\centerline{\resizebox{0.45\textwidth}{!}{%
  \includegraphics{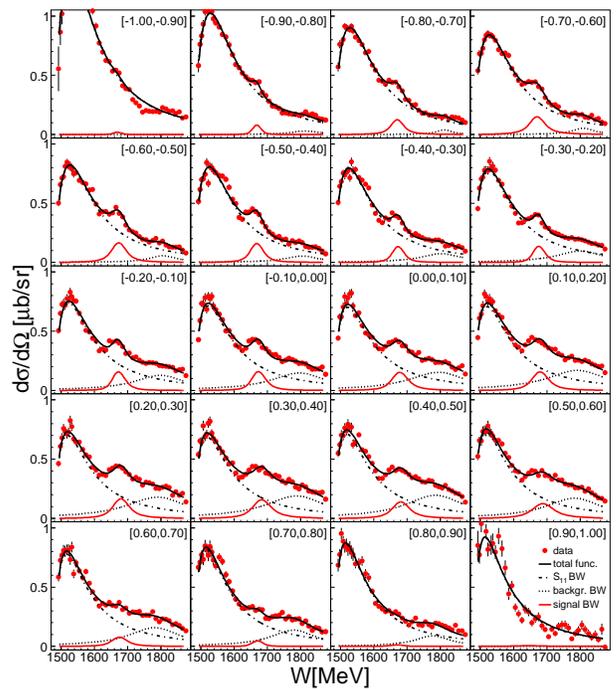}
}}
\caption{Excitation functions for $\gamma n\rightarrow n\eta$ for different ranges
of cos($\Theta_{\eta}^{\star}$) measured with the deuterium target. 
The fit curves are as in Fig.~\ref{fig:total}.
}
\label{fig:ang}       
\end{figure}

The structure in the neutron excitation functions is much more pronounced than for 
the Fermi smeared results as function of $E_{\gamma}$ shown in Fig.~\ref{fig:total}, top. 
The peak in the $\sigma_n/\sigma_p$ ratio appears at the same position for the results
from both nuclei and agrees also with the previous deuteron data \cite{Jaegle_11a}.
The data have been fitted with the ansatz from Ref. \cite{Jaegle_11a}, using 
a Breit-Wigner (BW) curve with energy dependent width for the S$_{11}$(1535) 
state (dash-dotted curves in Fig.~\ref{fig:total}), and two further Breit-Wigner curves, 
one as a phenomenological parameterization of all background contributions (dotted curve)
and one for the narrow structure (dashed curve). Note that the small structure below 1.6 GeV 
in the difference spectrum of data and background fit is an artifact because the 
simple ansatz for the fit curve does not exactly reproduce the lineshape in the high
energy tail of the S$_{11}$ resonance peak.
The fitted BW parameters of the narrow structure from the present experiments and from 
Jaegle et al. \cite{Jaegle_11a} are compared in Table~\ref{tab:bwfit}. 
They are in good agreement but represent only upper limits for the width of the structure which 
includes the contribution from the experimental resolution for $W$. 

The results from the deuterium target, which have smaller systematic uncertainties
due to the simple three-body final state, were analyzed in more detail. Excitation
functions for the $\eta\rightarrow 2\gamma$ and $\eta\rightarrow 6\gamma$
decays are shown separately in Fig.~\ref{fig:dexi}. For these data the experimental resolution
has been determined with a Monte Carlo simulation. The event generator produced the phase-space 
decay of a resonance at the position of the observed structure with zero intrinsic width 
($\delta$-function). The events were tracked through the detector system using the
Geant4 \cite{GEANT4} code and analyzed in the same way as the experimental data. 
The results are shown as black, dashed curves in Fig.~\ref{fig:dexi}, which have a width 
(FWHM) of $\approx$27~MeV, reflecting the experimental energy and angular resolution. 
Subsequently, the phenomenological fits described above were repeated but now the BW
curve for the narrow structure was numerically folded with the experimental resolution.
The result for the BW curve should then represent the intrinsic width of the structure
(solid, light blue curves in Fig.~\ref{fig:dexi}). The results of this analysis
are also shown in Table \ref{tab:bwfit}. 
The position of the structure is not influenced by the resolution. Also the signal strength
(which is proportional to the square root of the integral of the peak) is similar for the 
two analyses. But the widths are decreased to values around 30 MeV.   

\begin{table}[t]
\begin{center}
\caption{Fitted BW parameters (see text for description of fit ansatz
and Figs.~\ref{fig:total} (bottom) and \ref{fig:dexi} for fit curves)
of narrow structure. $W_R$ and $\Gamma$: position and width.
Electromagnetic coupling $A_{1/2}^n$ extracted under the assumption of an s-wave resonance. 
Values for width in brackets: fit with BW curve, 
without brackets: BW folded with experimental resolution. 
$^{*)}$ the intrinsic width for $^3$He is calculated assuming the experimental resolution
is the same as for the deuteron.
For Ref.~\cite{Jaegle_11a} (a) corresponds to the parameters given in the 
reference (analysis with cut on the spectator momentum) and (b) to an analysis without  
this cut.} 
\label{tab:bwfit}       
\begin{tabular}{|c|c|c|c|}
\hline\noalign{\smallskip}
& $W_R$ [MeV] & $\Gamma$ [MeV] & $\sqrt{b_{\eta}}A_{1/2}^n$ \\
& & & [10$^{-3}$GeV$^{-1/2}$] \\
 \hline
$^2$H, 2$\gamma$ \& 6$\gamma$ &  1670$\pm$1 & 29$\pm$3  (50$\pm$2)    &  12.3$\pm$0.8 \\ 
\hline
$^2$H, 2$\gamma$              &  1670$\pm$1 & 27$\pm$3  (50$\pm$3)    &  12.1$\pm$0.8 \\ 
\hline
$^2$H, 6$\gamma$              &  1669$\pm$1 & 30$\pm$5  (49$\pm$4)    &  12.9$\pm$0.8 \\ 
\hline
$^3$He 2$\gamma$ \& 6$\gamma$ &  1675$\pm$2  &  46$\pm$8$^{*)}$ (62$\pm$8) & 11.9$\pm$1.2\\
\hline
best estimate & 1670$\pm$5 & 30$\pm$15 & 12.3$\pm$0.8 \\
\hline
\hline
$^2$H \cite{Jaegle_11a}$^{(a)}$ & 1663$\pm$3 & (25$\pm$11) & 12.2$\pm$3 \\
$^2$H \cite{Jaegle_11a}$^{(b)}$ & 1673$\pm$4 & (54$\pm$16) & 16$\pm$3 \\
 \hline
\end{tabular}
\end{center}
\end{table}

For the $^3$He data the contribution of the experimental resolution to the effective
width is not exactly known due to the approximations made in the kinematic reconstruction.
In Table~\ref{tab:bwfit} we quote a value obtained under the assumption that the resolution
would be the same as for deuterium, i.e. the relative momentum of the spectator nucleons 
is negligible (which results in an upper limit for the width).

Finally, for the deuteron results, the dependence on the 
$\eta$ polar angle in the center-of-momentum frame was analyzed. The results are summarized in 
Fig.~\ref{fig:ang}, where excitation functions are shown for narrow bins of 
cos($\Theta_{\eta}^{\star}$). The strength of the structure is similar for a large range of 
polar angles, but it disappears towards extreme forward and backward angles. The results for 
the $^3$He measurement (not shown) behave similarly.

In conclusion, the present results demonstrate beyond any doubt the existence of a narrow 
structure in the excitation function of $\gamma n\rightarrow n\eta$. We estimate a position of 
$W$ = (1670$\pm$5)~MeV and an intrinsic width of $\Gamma$ = (30$\pm$15)~MeV.
These values are not statistically weighted averages of the entries from Table~\ref{tab:bwfit}
but reflect the range of observed variations, taking into account that the $^3$He results 
for the width are only upper limits. The central value reflects the deuterium results and
the uncertainties have been enlarged so that also the $^3$He results fall within the range.
When treated like an s-wave resonance the corresponding coupling 
strength $\sqrt{b_{\eta}}A_{1/2}^n$ is approximately (12.3$\pm$0.8) 10$^{-3}~$GeV$^{-1/2}$. 
If the structure actually corresponds to a nucleon resonance, it would have very unusual 
properties, but also other suggestions like the strangeness threshold effects 
discussed in \cite{Doering_10} or interference terms between different resonances 
\cite{Shklyar_07,Shyam_08,Anisovich_09} have been put forward. The precise results for 
the angular distribution of $\gamma n\rightarrow n\eta$ will allow stringent tests of 
models. Measurements of further observables, exploiting polarization degrees of freedom,
are already under way.

\vspace*{0.cm}
{\bf Acknowledgments}

We wish to acknowledge the outstanding support of the accelerator group 
and operators of MAMI. 
This work was supported by Schweizerischer Nationalfonds
(200020-132799,121781,117601,113511), Deutsche
Forschungsgemeinschaft (SFB 443, SFB/TR 16), DFG-RFBR (Grant No. 05-02-04014),
UK Science and Technology Facilities Council, (STFC 57071/1, 50727/1), 
European Community-Research Infrastructure Activity (FP6), the US DOE, US NSF and
NSERC (Canada).

\end{document}